\documentclass[12pt]{article}

\usepackage{multirow}
\usepackage[table,xcdraw,booktabs]{xcolor}
\usepackage{graphicx}
\usepackage{caption}
\usepackage{times}
\usepackage{booktabs}
\usepackage{amsfonts}
\usepackage{titlesec}
\usepackage{ulem}
\usepackage{natbib}

\begin{document} 

\title{Multi-task multi-station earthquake monitoring: An all-in-one seismic Phase picking, Location, and Association Network (PLAN)}


\author
{Xu Si$^{1}$, Xinming Wu$^{1\ast}$, Zefeng Li$^{1\ast}$, Shenghou Wang$^{2}$ and Jun Zhu$^{1}$\\
\normalsize{$^{1}$School of Earth and Space Sciences, University of Science and Technology of China,}\\
\normalsize{Hefei, China.}\\
\normalsize{$^{2}$Key Laboratory of Tectonics and Petroleum Resources of Ministry of Education,}\\
\normalsize{China University of Geosciences, Wuhan, China.}\\
\\
\normalsize{$^\ast$To whom correspondence should be addressed:}\\
\normalsize{E-mail: xinmwu@ustc.edu.cn and zefengli@ustc.edu.cn}
}
\date{}






\maketitle

\begin{abstract}
Earthquake monitoring is vital for understanding the physics of earthquakes and assessing seismic hazards. 
A standard monitoring workflow includes the interrelated and interdependent tasks of phase picking, association, and location. 
Although deep learning methods have been successfully applied to earthquake monitoring, 
they mostly address the tasks separately and ignore the geographic relationships among stations. 
Here, we propose a graph neural network that operates directly on multi-station seismic data and achieves simultaneous phase picking, association, and location. 
Particularly, the inter-station and inter-task physical relationships are informed in the network architecture to promote accuracy, interpretability, and physical consistency among cross-station and cross-task predictions. 
When applied to data from the Ridgecrest region and Japan regions, this method showed superior performance over previous deep learning-based phase-picking and localization methods. 
Overall, our study provides for the first time a prototype self-consistent all-in-one system of simultaneous seismic phase picking, association, and location, which has the potential for next-generation autonomous earthquake monitoring.
\end{abstract}


\section*{Introduction}

Earthquake monitoring is one of the most fundamental operations in seismology. 
A standard earthquake monitoring workflow involves a series of steps to detect and characterize earthquakes,
including phase picking, association, and event location~\citep{beroza2021machine,mousavi2022deep,zhu2022end}.
Phase picking, a conceptually simple task which is akin to detection problems in computer vision, has recently been improved through deep learning~\citep{zhu2022end,ross2018generalized,ross2018p,zhu19phasenet,mousavi2019cred,
zhu2019deep,pardo2019seismic,wang2019deep,liu2020rapid,mousavi2020earthquake,yang2021simultaneous,yano2021graph,
zhu2022ustc,bilal2022early,feng2022edgephase,munchmeyer2022picker}, where convolutional neural networks (CNNs)~\citep{krizhevsky2017imagenet} are typically used.
After the phase picking, traditional~\citep{arora2013net,gibbons2016iterative,zhang2019rapid,zhu2022earthquake} 
and deep-learning-based~\citep{ross2019phaselink,mcbrearty2019pairwise,yu2022fastlink} phase association 
algorithms have been used to link seismic phases at multiple stations from the same events.
Finally, location algorithms~\citep{bakun1997estimating} utilize the associated phases to obtain the 
earthquake hypocenters,
although some deep-learning-based methods directly process raw data to locate earthquakes~\citep{zhang2014real,devries2018deep,perol2018convolutional,lomax2019investigation,mousavi2019bayesian,zhang2020locating,munchmeyer2021earthquake}.

These three tasks (phase picking, association, and location) are closely interdependent.
The accuracy of multi-station phase picking affects the accuracy of association and location. 
Conversely, association and location impose constraints on multi-station phase picking. 
Additionally, phase picking with multi-station data can further utilize the geographic relationships and waveform similarities among multiple stations. 
To achieve more efficient and accurate earthquake monitoring, a suitable earthquake monitoring workflow should impose inter-task and inter-station constraints and preferrably perform all three tasks simultaneously at all stations. 
However, most existing earthquake monitoring methods perform phase picking, association, and earthquake location separately. 
In addition, most of the current phase-picking methods process seismic data on a station-by-station basis.
While some recent graph-based approaches~\citep{mcbrearty2019earthquake,van2020automated, mcbrearty2022earthquake1, mcbrearty2023earthquake,zhang2022spatio} have demonstrated the ability to handle irregularly spaced stations for phase association and event location, it remains a challenging task to develop a method that effectively leverages inter-task and inter-station constraints, and ideally performs all three tasks simultaneously.	

Here, we propose an all-in-one earthquake monitoring system called seismic Phase picking, Location, and Association Network (PLAN) that achieves for the first time the simultaneous implementation of the three tasks with multi-station data and inter-task constraints. 
PLAN consists of four interdependent neural network modules. Specifically, the first module of waveform feature extraction utilizes an encorder-decoder architecture to extract relevant features from multi-station seismic data. 
The second module of earthquake location encodes station locations (i.e., longitude, latitude, and elevation) and merges them with waveform features from the first module to predict the earthquake depth and epicentral distance for each station. 
The third module of phase association utilizes the predicted earthquake location information to estimate the time shifts required to align multi-station waveform features. 
Finally, the fourth module of phase picking aggregates the aligned features for simultaneous multi-station phase picking. 
We applied PLAN in the Ridgecrest and Japan regions and compared its efficiency and accuracy with that of state-of-the-art phase-picking and event location methods, demonstrating the merits of inter-station and inter-task constraints for accurate earthquake monitoring.

\section*{Results}
\subsection*{Network architecture}

\begin{figure}[b!]%
\centering
\includegraphics[width=\textwidth]{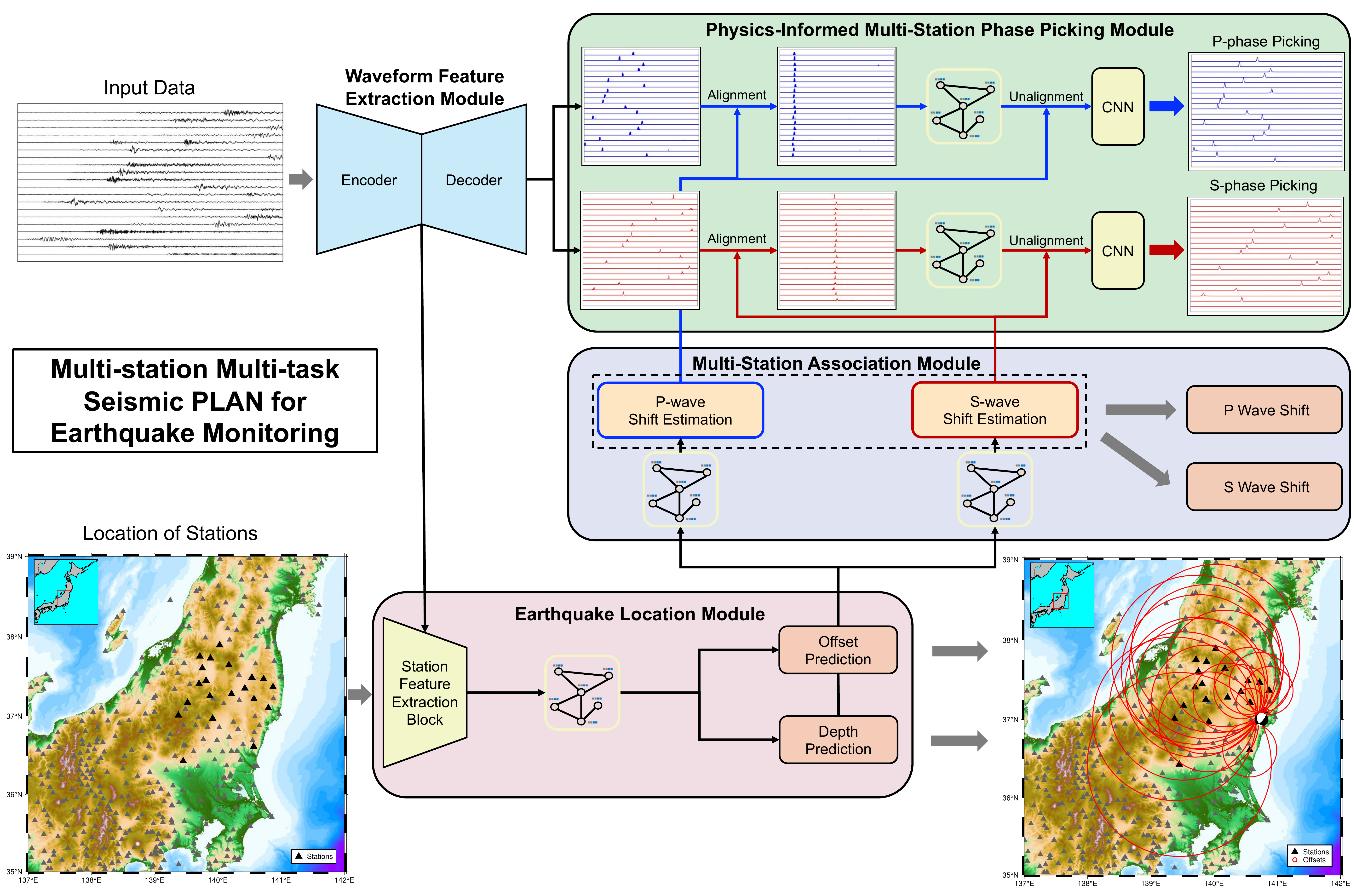}
\captionsetup{font={small}}
\caption{
The flowchart of the proposed multi-task and multi-station PLAN for earthquake monitoring.
The input data for the model comprise seismic waveforms recorded by multiple stations and 
the locations of these stations. 
PLAN consists of four sub-modules: waveform feature extraction encoder and decoder, 
an earthquake location module, a multi-station association module and a physics-informed multi-station phase-picking module. All of these sub-modules are optimized simultaneously and constrained by each other during training to improve performance in earthquake detection, association, and location.
}
\label{fig1}
\end{figure}

The proposed multi-station multi-task PLAN (Fig. \ref{fig1}) employs a Graph neural network 
(GNN)~\citep{kipf2016semi} as the backbone to integrate the four functional modules of waveform feature extraction, earthquake location, multi-station association, and a physics-informed multi-station phase picking (further details are provided in the Methods section). Compared with CNN, GNN is naturally suited for handling seismic data acquired from irregularly spaced stations whose quantity and location may vary~\citep{mcbrearty2023earthquake}. In constructing our GNN, we define the graph nodes with seismic stations and define the feature vector for each node with the location and recoded seismic signal of the corresponding station. All nodes (corresponding to the seismic stations) are linked together and the linking weights are learnt during training to infer the relationships among the stations. 
We construct the GNN layers with TransformGConvs~\citep{shi2020masked}, which are designed based on an attention mechanism~\citep{vaswani2017attention} to learn the dynamically linking weights among different stations. (details about TransformGConvs are provided in the Methods section). In addition, the graph nodes are not fixed so that the GNN could be adapted to variations in the station number and location.

Raw three-component seismic signals are input into the graph nodes. The front-end waveform feature extraction module, constructed as an encoder-decoder CNN and shared among the nodes, extracts their corresponding key features. 
The station feature extraction block, constructed as two MLPs and shared among the nodes, extract geographic features from the normalized input longitudes, latitudes, and elevations of the stations. 
The earthquake location module then concatenates the extracted waveform and geographic features and employs multiple TransformGConvs to aggregate these features from multiple nodes to predict the event depth and station-event offset. The predicted offsets and depth are further used to determine the event location by a triangulation algorithm~\citep{yu1996slip}. 
In the earthquake location module, we do not directly predict the event location but the station-event offsets and depth instead, because we will use them as input into a followed multi-station association module to estimate the time shifts needed to align the P and S arrivals.

This multi-station association module plays a key role in bridging the tasks of earthquake location and multi-station phase picking and introduces physical constraints between the two tasks. 
Prior to aggregating the waveform features from different stations for multi-station phase picking, 
the features corresponding to the same earthquake are required to be initially aligned or associated; otherwise, aggregation of unaligned features could mutually interfere and ultimately degrade the picking results.
The multi-station phase-picking module includes a non-trainable physical layer, implemented with the Pytorch~\citep{paszke2019pytorch} roll function, to shift and align the waveform features (from the decoder of the waveform feature extraction module) using the time shifts. 
Subsequently, multiple TransformGConvs in the phase-picking module aggregate the aligned waveform features to enhance the phase-picking features in the aligned space. 
Eventually, another physical layer unshifts the aggregated features back to the original space, followed by two convolutional layers to obtain the P/S-wave picks at all the stations.

Three regression loss functions are defined for the three modules of phase picking, association, and earthquake location and then combined to jointly train the entire network. 
Because all the modules are interconnected within the entire network, the training process finds an optimal network that could perform all the tasks both accurately and consistently. Moreover, after training, the multi-station association module could be detached from the network and utilized to calculate the S-P differential travel time with inputs of offsets and event depth. Further details on this module are provided in the section titled “Multi-station association module”.

\subsection*{Data preparation}

We tested the proposed PLAN in two regions of Ridgecrest and Japan. For the Ridgecrest region (Fig. \ref{fig2}A), seismic recordings from 16 California Integrated Seismic Network stations within an epicentral distance of \textless~80 km were collected from July 4, 2019, to October 4, 2019, for a total of more than 71,000 M~\textgreater~0 earthquakes. The data for Japan (Fig. \ref{fig3}A) included M~\textgreater~2 earthquakes that occurred between January 1, 2011, and December 31, 2011, including the Mw 9.1 Tohoku sequence. 
We collected the 3-component High Sensitivity Seismograph Network (NIED Hi-net)~\citep{obara2005densely,aoi2020mowlas} data from over 35,000 events. Subsequently, the data were randomly divided into training, validation, and test sets (85\%, 5\%, and 10\%, respectively) in both regions. 

The number of stations corresponding to each event in the training samples varied, and the trained network can flexibly handle situations where the number of stations changes in actual data.
Further, the distributions of the number of stations per event in the training and test sets were balanced. 
The results for the test sets in the two regions are presented in Figs. \ref{fig5} and \ref{fig6}, respectively. 
To accommodate different range scales in the two study regions, we used different window lengths in two regions (30.72 s for Ridgecrest and 61.44 s for Japan) with the same sampling frequency (100 Hz).

To ensure a fair comparison with the existing phase-picking methods, we followed the same data preprocessing procedures 
used in previous studies~\citep{zhu19phasenet,mousavi2020earthquake}. 
This involved normalizing the data by removing the mean and dividing by the standard deviation,   
and using a Gaussian-shaped target function as training labels for the P/S-phase arrival times. 
The use of Gaussian-shaped targets is effective in phase picking ~\citep{zhu19phasenet,mousavi2020earthquake}, 
and in our study, the probabilities of P wave and S wave arrival times were set to 1 at the first arrival time 
and decreased to 0 within a 20 sample window before and after each phase arrival.

\subsection*{Application to the 2019 Ridgecrest sequence}\label{subsec3}

\begin{figure}[!h]%
\centering
\includegraphics[width=\textwidth]{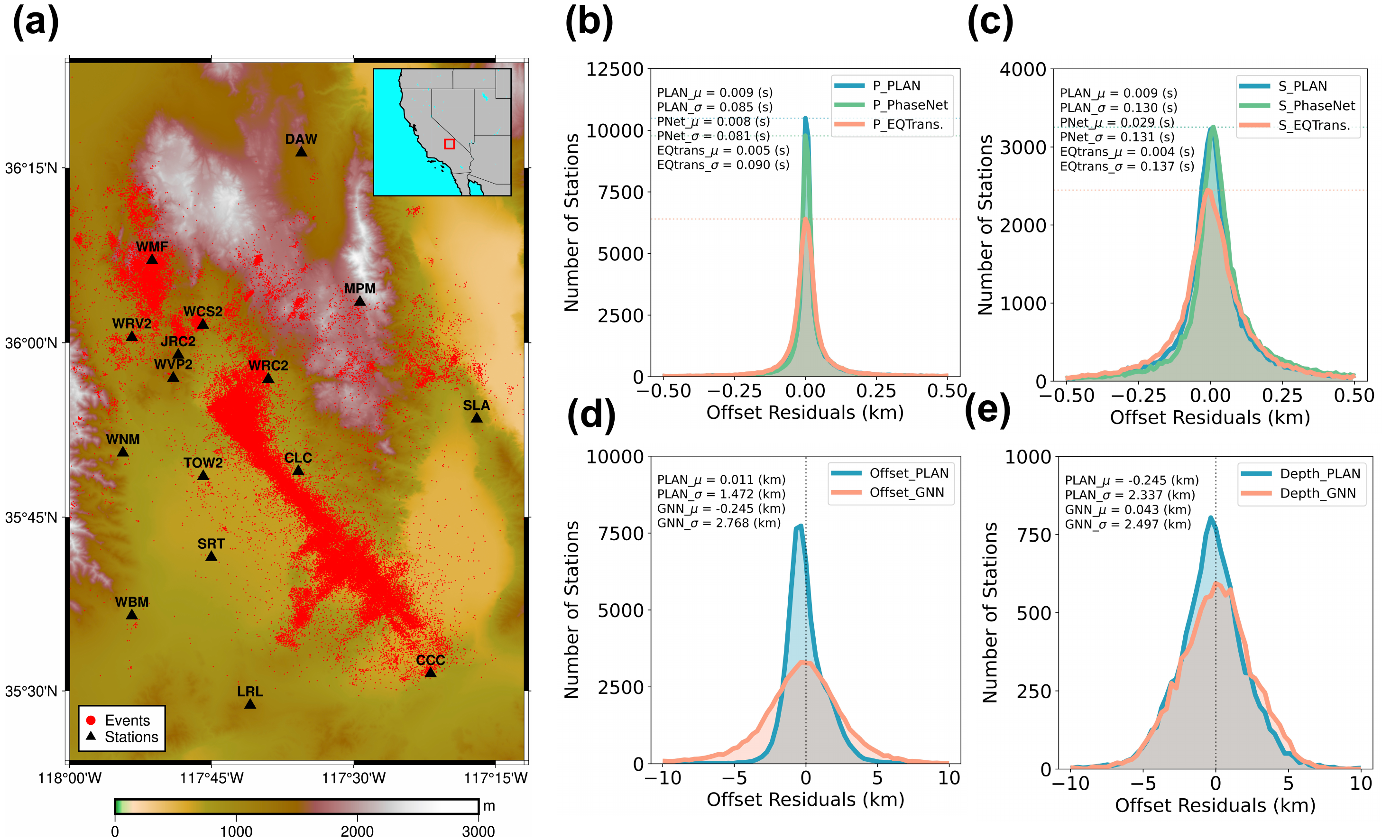}
\captionsetup{font={small}}
\caption{
Distributions of phase picking and location residuals in Ridgecrest region.
(\textbf{A}) Distribution of 16 stations (black triangles) and training event locations (red circles) used in our study,
where the events occurred between July 4, 2019 and October 4, 2019.
(\textbf{B} and \textbf{C}) The results of P-wave and S-wave arrival time residual, respectively.
The blue, green, and orange lines in (B and C)
represent the arrival time residuals for PLAN, PhaseNet, and 
EQTransform, respectively. 
The proposed method yields the most accurate results in P/S-wave picking.
(\textbf{D} and \textbf{E}) The offset and depth residuals between model predictions and Southern California Seismic Network (SCSN) catalog of the located events.
Regardless of the offset or depth, the residual distribution of PLAN (blue line) is more concentrated at zero than that of Aggregated-GNN (orange line).
}\label{fig2}
\end{figure}

We compared the performance of PLAN with that of other established deep learning methods for earthquake phase picking (PhaseNet~\citep{zhu19phasenet} and EQTransformer~\citep{mousavi2020earthquake}) and location (Aggreated-GNN~\citep{van2020automated}). 
 All of the methods were retrained on the same training set and evaluated on a common test set.
 As shown in the Ridgecrest application in Fig. \ref{fig2}B-\ref{fig2}C, the performance of PLAN in phase picking was superior to that of the other two deep learning-based methods. Specifically, the residual distribution of the P-wave picks for PLAN was more concentrated than that of the other methods, indicating a higher overall accuracy. 
 For S-wave picks, PLAN performed significantly better than EQTransformer because the distribution of PLAN was narrower whereas the difference in performance between PLAN and PhaseNet was relatively minor.

In terms of localization, our method (PLAN) outperformed the other approaches (Fig.~\ref{fig2}D-\ref{fig2}E; Table~\ref{tab1}). 
The distribution of PLAN was notably more concentrated than that of Aggregated-GNN, particularly in terms of offset prediction. 
To further demonstrate the effectiveness of TransformGConv, we replaced the TransformGConv layers (Fig. S1)
with GCN~\citep{kipf2016semi}, SAGE~\citep{hamilton2017inductive},
and GATv2~\citep{brody2021attentive} respectively.
Among the various methods compared, our method yielded the best results in terms of offset, with an average error of 1.09 km and a standard deviation of 1.41 km. Furthermore, PLAN also outperformed Aggregated-GNN in terms of depth localization, regardless of whether it was based on GCN, GATv2, or TransformGConv. These results demonstrated the superiority of the proposed PLAN in location estimations.

Furthermore, we used three metrics of mPrecision, mRecall, and mF1 (described in the Methods section), to quantitatively evaluate the performance of the five methods (Table~\ref{tab2}). 
In five of the six metric scores for the P-wave and S-wave picking results, our attention mechanism-based GNN method outperformed the other methods. 
The only exception was the mPrecision metric of P-wave picking, where the EQTransformer showed slightly higher scores than PLAN. 
Notably, even the simplified version of the multi-station phase-picking method, such as the SAGE-based PLAN, outperformed both the single station-based picking methods of EQTransformer and PhaseNet in mF1 scores for S-wave picking. This indicated that the phase-picking accuracy is significantly improved by multi-station picking, which effectively utilizes inter-station contextual information. We not only adjusted the time threshold while maintaining a constant picking probability for evaluation but also fixed the time threshold and changed the  probability of picking to calculate and plot the precision-recall curves for four models (Fig. S2). Consistent with the aforementioned results, the PLAN model, based on TransformGConv, demonstrated superior performance in terms of F1, encompassing both P-wave and S-wave.


\begin{table}[htb!]
\begin{center}
\begin{minipage}{\textwidth}
\caption{Location Performance in Ridgecrest and Japan regions.}\label{tab1}
\begin{tabular*}{\textwidth}{@{\extracolsep{\fill}}lcccccc@{\extracolsep{\fill}}}
\toprule	
& & & \multicolumn{2}{@{}c@{}}{Offset MAE (km)} & \multicolumn{2}{@{}c@{}}{Depth MAE (km)} \\ \cline{4-5} \cline{6-7}
Region & \multicolumn{2}{@{}c@{}}{Method} & Mean & Std & Mean & Std  \\
\midrule
& \multicolumn{2}{@{}c@{}}{Aggregated-GNN} & 2.30 & 2.30 & 1.98 & 1.55 \\
&&GCN  & 8.96 & 6.94 & 1.43 & 1.34  \\
&&GATv2  & 8.95 & 6.90  & \color[HTML]{FE0000} \textbf{1.42}  & 1.33 \\
&&{SAGE}  &  1.28 & 2.09 & 1.68 & 1.42 \\
\multirow{-5}{*}{Ridgecrest} & \multirow{-4}{*}{PLAN} &Trans  &  \color[HTML]{FE0000} \textbf{1.09} &  \color[HTML]{FE0000}  \textbf{1.41}  &  1.43  &  \color[HTML]{FE0000} \textbf{1.33} \\
\midrule
& \multicolumn{2}{@{}c@{}}{Aggregated-GNN} & 27.92 & 26.78 &  10.30 & 8.95 \\
 &&GCN  & 21.30 & 16.47 & 13.50 & 9.95 \\
 &&GATv2  & 10.79 & 10.82  & \color[HTML]{FE0000} \textbf{5.59}  & \color[HTML]{FE0000} \textbf{6.86} \\
 &&SAGE  &  4.87 & 6.11 & 12.85 & 9.41 \\
 \multirow{-5}{*}{Hinet} & \multirow{-4}{*}{PLAN}  &Trans&  \color[HTML]{FE0000} \textbf{4.81} &  \color[HTML]{FE0000}  \textbf{5.83}  &  10.62  &   8.89 \\
\bottomrule
\end{tabular*}
\footnotetext{Note: Red and bold values represent the best performance.}
\end{minipage}
\end{center}
\end{table}

\begin{table}[htb!]
\begin{center}
\begin{minipage}{\textwidth}
\caption{
Detection Performance in Ridgecrest region.
}\label{tab2}
\begin{tabular*}{\textwidth}{@{\extracolsep{\fill}}lcccccc@{\extracolsep{\fill}}}
\toprule
& \multicolumn{3}{@{}c@{}}{P Picking Metrics} & \multicolumn{3}{@{}c@{}}{S Picking Metrics} \\\cmidrule{2-4}\cmidrule{5-7}%
Method & mPrecision & mRecall & mF1 & mPrecision & mRecall & mF1 \\
\midrule
PhaseNet  & 94.83 & 92.78 & 93.79 & 84.50 & 80.65 & 82.53\\
EQtransformer  & \color[HTML]{FE0000} \textbf{95.43} & 91.17  & 93.25  & 86.77 & 78.21 & 82.27\\
\midrule
PLAN-GATv2  & 95.05 & 93.02  & 94.03  & 85.55 & 80.49 & 82.95\\
PLAN-SAGE  &  94.99 & 93.07 & 94.02 & 85.65 & 81.48 & 83.51\\
PLAN-Trans  & 94.65 & \color[HTML]{FE0000} \textbf{94.90}  & \color[HTML]{FE0000} \textbf{94.77}  & \color[HTML]{FE0000} \textbf{86.88} & \color[HTML]{FE0000} \textbf{84.94} & \color[HTML]{FE0000} \textbf{85.90}\\
\bottomrule
\end{tabular*}
\footnotetext{
Note: The picking probability is 0.3. Red and bold values represent the best performance.}
\end{minipage}
\end{center}
\end{table}

\subsection*{Application to seismicity in Japan}\label{subsec4}

\begin{figure}[!h]%
\centering
\includegraphics[width=\textwidth]{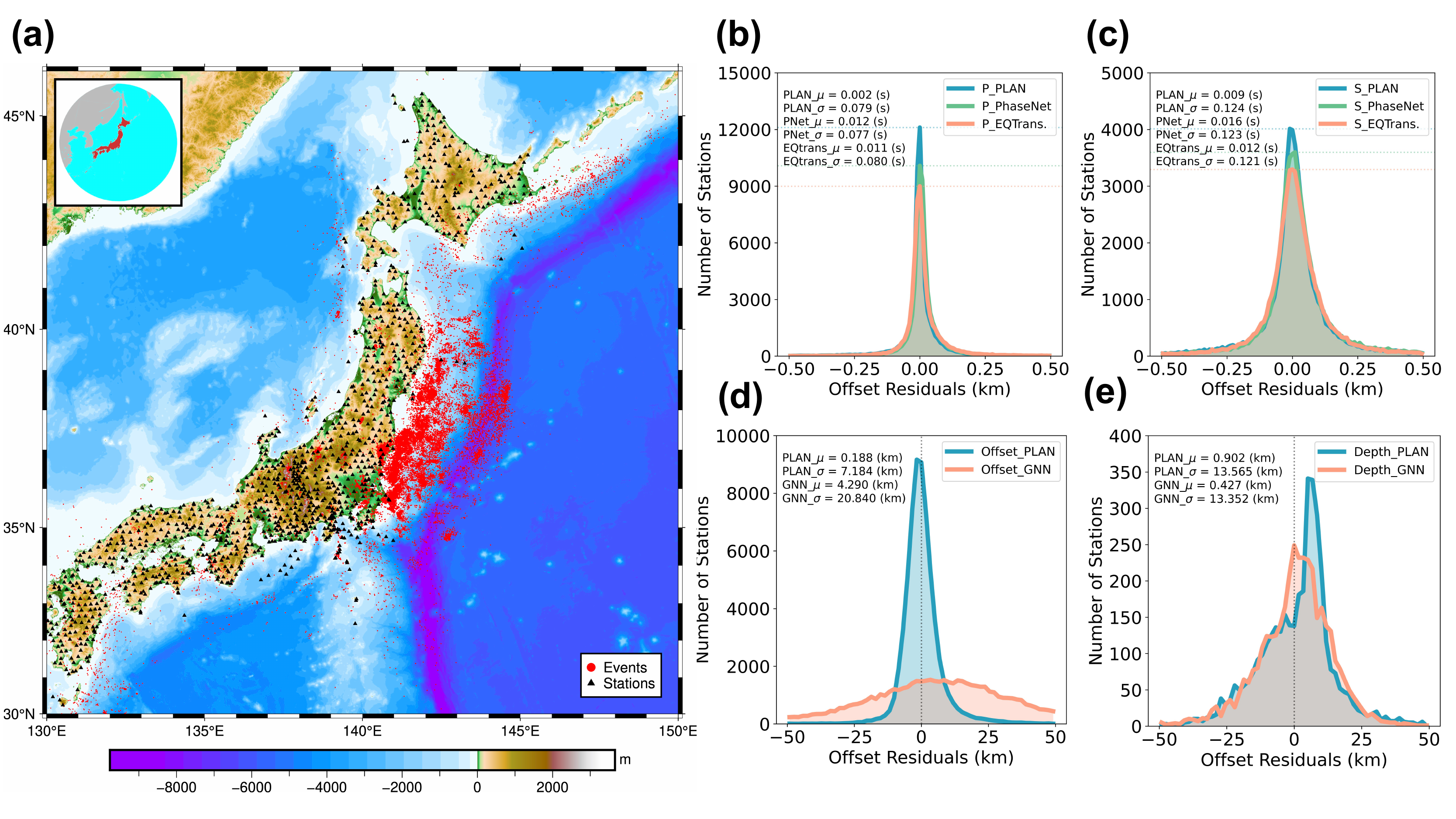}
\caption{Distributions of phase picking and location residuals in Japan.
(\textbf{A}) Distribution of stations (black triangles) and training event locations (red circles) used in our study,
where the events occurred between January 1, 2011 and December 31, 2011.
(\textbf{B} and \textbf{C}) Similar to Fig. \ref{fig2}, (B and C) show the results
of P-wave and S-wave arrival time residuals, respectively. 
(\textbf{D} and \textbf{C}) The offset and depth residuals 
between model predictions and the Japan Meteorological Agency (JMA)
catalog of located events, respectively.
}\label{fig3}
\end{figure}

We retrained all the methods on the Japan training set for the evaluation.
Compared to its performance in the Ridgecrest region, PLAN exhibited an even better performance in Japan (Fig. \ref{fig3}B-\ref{fig3}C).
Further, PLAN demonstrated a remarkably better performance than PhaseNet and EQTransformer for both P- and S-wave picks.
The offset predicted by PLAN was notably more accurate than that predicted by Aggregated-GNN, with higher and narrower residuals (Fig. \ref{fig3}D-\ref{fig3}E).
In terms of depth estimation, although PLAN maintained a narrower residual distribution, 
the highest point of the distribution was shifted systematically, compared with the Aggregated-GNN method.
Table \ref{tab1} presents the comprehensive quantitative comparison of the results.
Although the TransformGConv-based PLAN method did not demonstrate particular superiority in depth estimation,
it excelled in offset estimation. 
Further, the GATv2-based PLAN showed the lowest depth error, indicating potential improvement of localization capabilities of the proposed PLAN.

\begin{table}[b!]
\begin{center}
\begin{minipage}{\textwidth}
\caption{Detection Performance in Japan.}\label{tab3}
\begin{tabular*}{\textwidth}{@{\extracolsep{\fill}}lcccccc@{\extracolsep{\fill}}}
\toprule
& \multicolumn{3}{@{}c@{}}{P Picking Metrics} & \multicolumn{3}{@{}c@{}}{S Picking Metrics} \\\cmidrule{2-4}\cmidrule{5-7}%
Method & mPrecision & mRecall & mF1 & mPrecision & mRecall & mF1 \\
\midrule
PhaseNet  & 94.87 & 94.97 & 94.92 & 88.26 & 84.38 & 86.28\\
EQtransformer  & \color[HTML]{FE0000} \textbf{95.91} & 94.79  & 95.35  & \color[HTML]{FE0000} \textbf{89.08} & 84.40 & 86.68\\
\midrule
PLAN-GATv2  & 95.14 & 93.81  & 94.47  & 88.35 & 81.87 & 84.98\\
PLAN-SAGE  &  95.65 & 94.90 & 95.27 & 88.45 & 84.19 & 86.27\\
PLAN-Trans  & 95.79 & \color[HTML]{FE0000} \textbf{95.14}  & \color[HTML]{FE0000} \textbf{95.46}  &  88.41 &   \color[HTML]{FE0000} \textbf{85.09} & \color[HTML]{FE0000} \textbf{86.72}\\
\bottomrule
\end{tabular*}
\footnotetext{
Note: The picking probability is 0.3. Red and bold values represent the best performance.}
\end{minipage}
\end{center}
\end{table}

Similar to the Ridgecrest example, we assessed the phase-picking performance of various models applied to the test data from Japan using mPrecision, mRecall, and mF1 metrics (Table \ref{tab3}) and precision-recall curves (Fig. S3). The TransformGConv-based PLAN model achieved  superior results in terms of mRecall (95.14 for P-waves and 85.09 for S-waves) and mF1 (95.46 for P-waves and 86.72 for S-waves), whereas EQTransformer performed best in terms of mPrecision of P-waves and S-waves. 
TransformGConv-based PLAN demonstrated high mRecall scores, indicating that a large proportion of the samples containing P/S-waves were correctly detected. 
However, this was achieved at the expense of a slightly lower mPrecision compared to that of the EQTransformer, with some non-P/S-waves incorrectly classified as P/S-waves.
The mF1 score provided a more comprehensive evaluation of the model performance, considering both the reduction in missed detections and the increase in correct detections. 
In this context, TransformGConv-based PLAN had the highest F1 score, indicating that it effectively reduced the missed detections of P/S-waves and increased the proportion of correct detections.

\subsection*{Multi-station association module}\label{subsec5}

To simultaneously pick P/S-waves from multiple stations, PLAN utilizes GNNs, which typically aggregates raw signals received at different stations. Feature aggregation across multiple stations introduces inter-station constraints and enhances the features at each station. However, because of different travel times of the same source across different stations, directly aggregating the signals from multiple stations would deteriorate multi-station picking. 

To address this issue, the proposed method employes a multi-station association module to estimate the time shifts as illustrated in Fig. \ref{fig1}. 
The input to this module is the offset of each station with respect to the event and its depth. 
The module output is the corrected time shifts of the P/S-wave for each station. 
To align the P/S-wave features, the correction criteria for the P/S-wave features were set at the 10 and 15 s in the Ridgecrest region and at the 20 and 32 s in Japan.
Using these criteria, the multi-station association module was trained to estimate the corrections of P/S-wave for each station. 
These corrections are then used to align the waveform features, enabling the graph convolution to aggregate the features in a temporally aligned space. 
Consequently, the method could enhance or compensate for the features at each station by fusing the aligned features from other stations, allowing simultaneous and accurate multi-station picking.

\begin{figure}[htb!]%
\centering
\includegraphics[width=\textwidth]{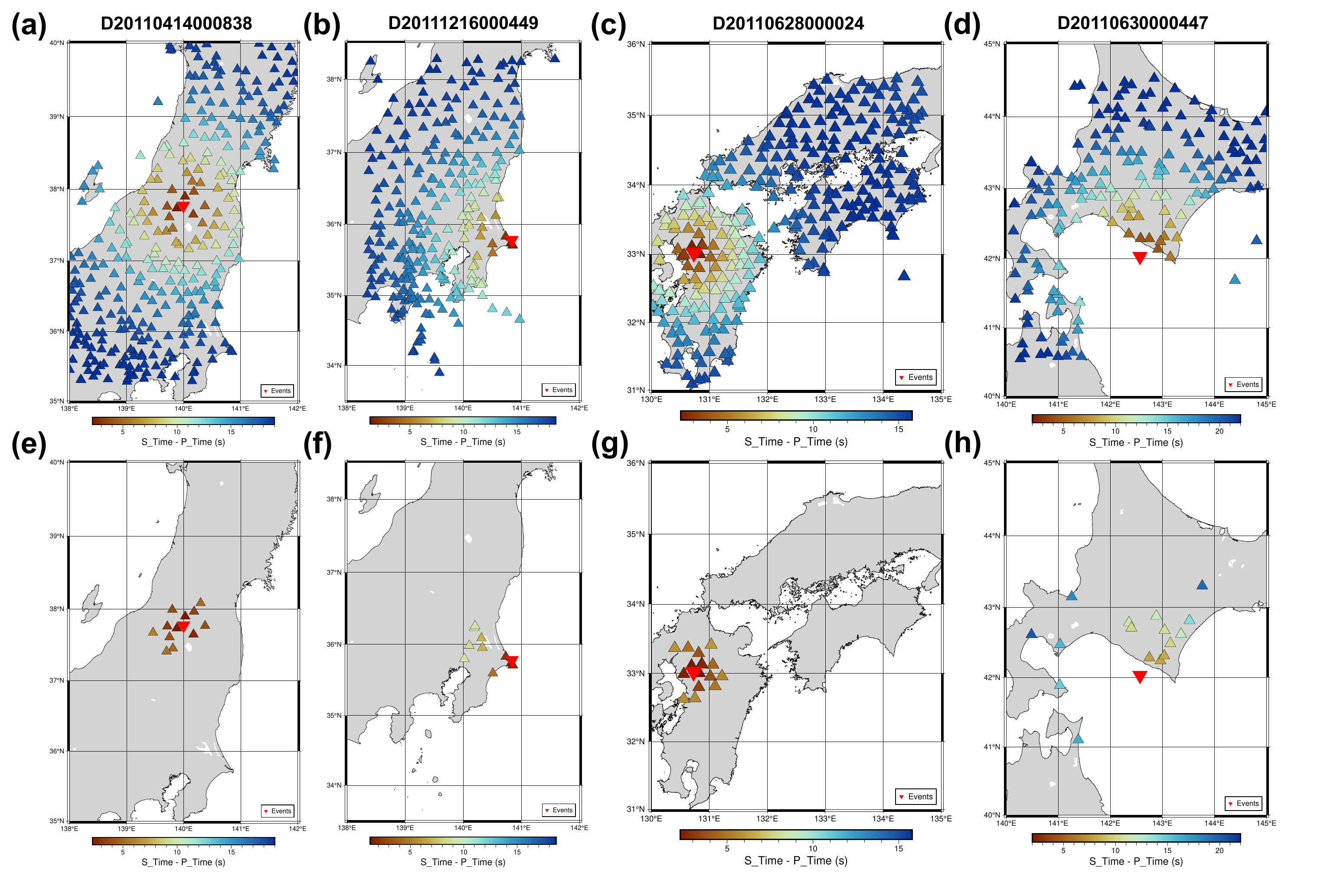}
\caption{
Arrival time residuals of P/S-waves at various stations in the test 
dataset of Japan. 
(\textbf{A} to \textbf{D}) Arrival time 
differences for each event calculated by multi-station association module, respectively, while
(\textbf{E} to \textbf{H}) show 
manually picked arrival time differences for the same events. 
The numbers on the top of the figures represent the identification numbers of the events in the Japan Meteorological Agency (JMA) catalog. 
Above results indicate that the S-P differential arrival time predicted by our multi-station association module are consistent with the manually picking differences. They also demonstrate that our module could calculate the arrival time difference for a larger number of stations.
}\label{fig4}
\end{figure}

The multi-station association module can be utilized independently after training. It converts the distance and depth information into arrival information and calculates the S-P differential travel time~\citep{crotwell1999taup}. 
Figure \ref{fig4} illustrates the arrival time differences of the P/S-waves at various stations in Japan. 
Although the training process utilizes a maximum of 37 stations for a single event, 
the module can be adapted to cases with any number of stations (e.g., hundreds of stations shown in Fig. \ref{fig4}A - Fig. \ref{fig4}D) 
to estimate the P/S-wave arrival time differences for all stations.
These results indicated that the module effectively enforced physical constraints based on time shifts within the overall network.

\section*{Discussion}

\begin{figure}[h!]%
\centering
\includegraphics[width=\textwidth]{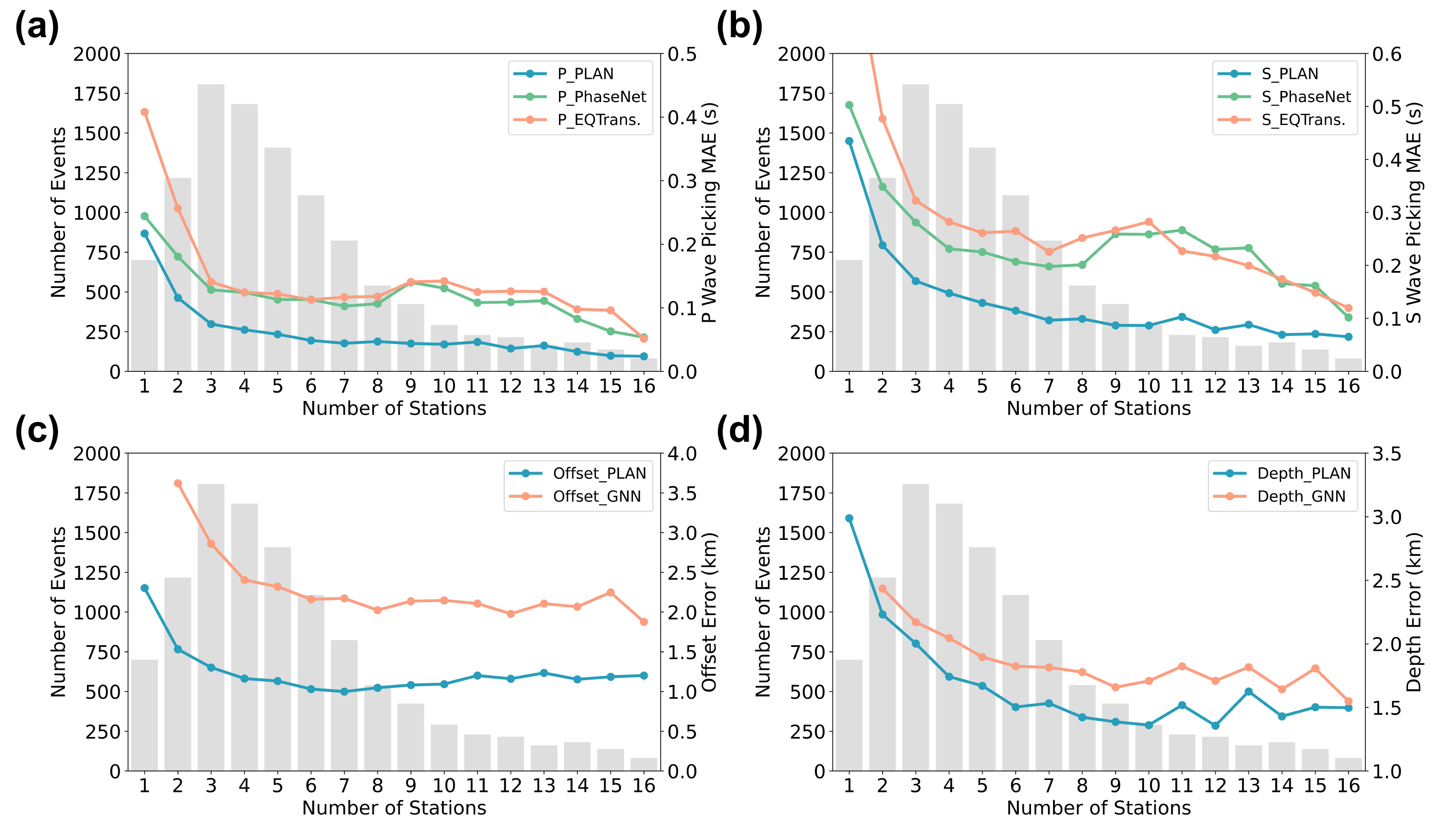}
\caption{
Comparison of prediction results using different numbers of stations in the Ridgecrest region.
(\textbf{A} to \textbf{D}) The distributions of prediction errors for P-wave, S-wave, offset, and depth, respectively.
The x-axis represents the number of stations, the primary y-axis
denotes the number of events recorded by a specific number 
of stations, and the secondary y-axis represents the prediction 
errors for phase picking and event localization.
Note that for phase picking, prediction residuals of PLAN (blue curves) decrease evidently as the number of stations increases. 
Moreover, the location errors of PLAN are significantly smaller than those of 
the Aggregated-GNN method.
}\label{fig5}
\end{figure}

PLAN is scalable for accommodating various numbers of stations per event. 
As PLAN is a network level picking and location model, we further investigated the effect of different numbers of stations on the network performance for phase picking and earthquake location with a test set of the Ridgecrest region (Fig. \ref{fig5}). 
We calculated the P- and S-wave picking residuals of the three different methods relative to the manual picking results, respectively (Fig. \ref{fig5}A-\ref{fig5}B).
The residuals of the single-station-based picking methods, PhaseNet and EQTransformer, exhibited oscillations for samples with station numbers 3-13 as the number of stations increased. 
Contrastingly, the residuals of our simultaneous multi-station picking method, PLAN, exhibited a significant residual decrease as the number of stations increased. 
Although the prediction residuals of the single-station-based methods should not be significantly associated with the number of stations, their prediction residuals still decreased when the number of stations was 13-16. 
This was probably because the events recorded by more stations tended to be larger and easier to pick.

A comparison of the distribution of prediction errors for earthquake offsets and depths with respect to the number of stations indicated that the errors in PLAN were significantly smaller than those in the Aggregated-GNN method (Fig. \ref{fig5}C-\ref{fig5}D). 
However, the errors in offset prediction did not exhibit a significant decrease with an increase in the number of stations. 
This was likely because a large number of stations would include more distant ones that tended to have large offset prediction errors. 
As the offset error metric is defined as the average value acquired from multiple stations, an increase in the number of stations can lead to a slightly higher average error for a single event.

\begin{figure}[b!]%
\centering
\includegraphics[width=\textwidth]{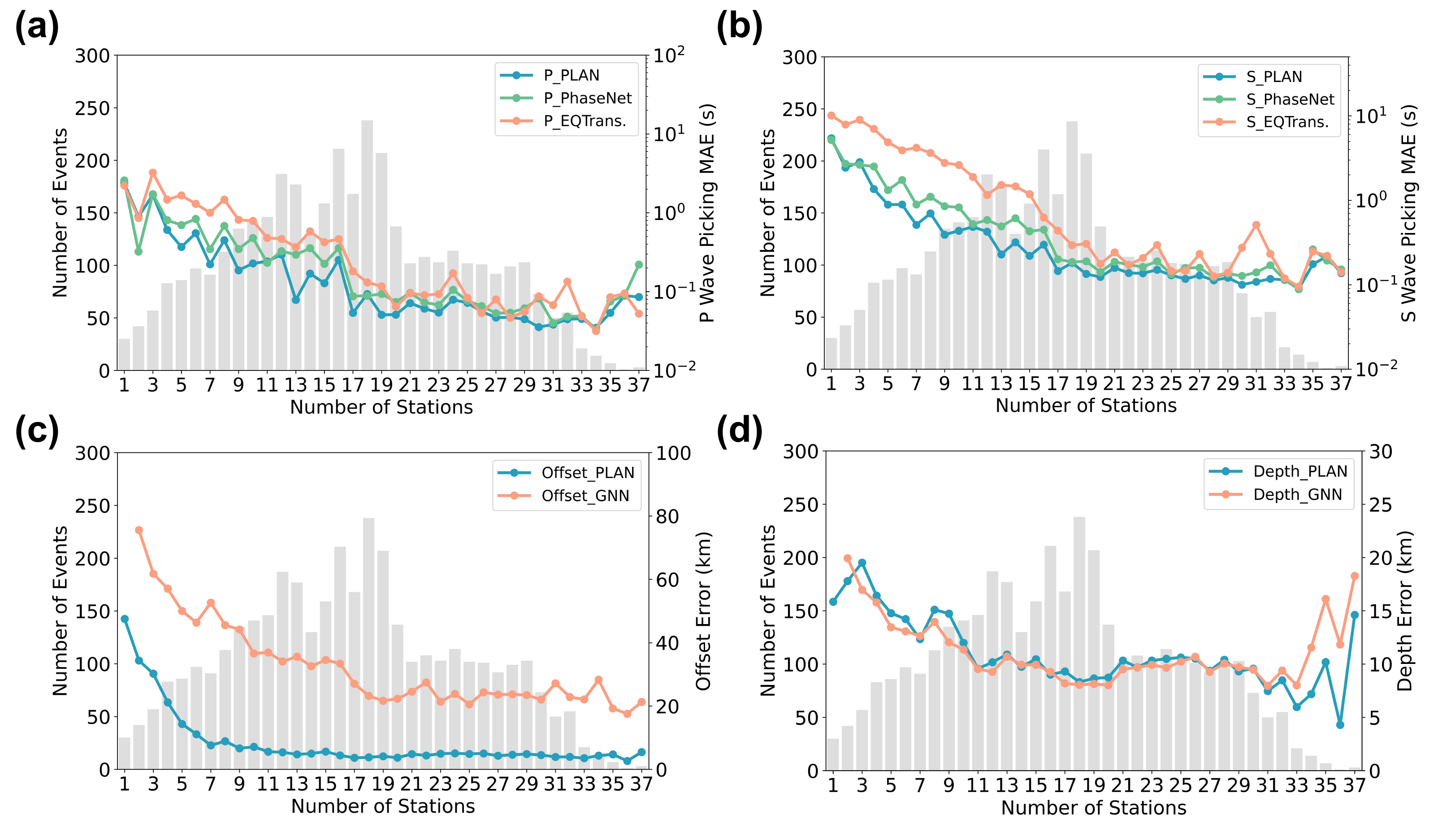}
\caption{
Comparison of prediction results using different number of stations in Japan. (\textbf{A} to \textbf{D}) The four distributions are similar to those described in Fig. \ref{fig5} and the only difference is that we have used logarithmic coordinates for the secondary y-axis in (A) and (B).
}\label{fig6}
\end{figure}

Furthermore, the statistical results for Japan (Fig. \ref{fig6}) 
were similar to those for the Ridgecrest region, with the PLAN method exhibiting smaller phase-picking errors than EQTransformer and PhaseNet, especially for S-wave picking. 
In addition, as the number of stations increased, the offset prediction error of PLAN became significantly smaller than that of the Aggregated-GNN. 

The ability of our network to handle varying numbers of stations can be attributed to the multi-station association module,
which can be separated from the entire network 
and utilized in a manner similar to the Taup algorithm for estimating the arrival time of earthquakes at stations.
Differing from the Taup algorithm, our association module does not depend on an input velocity model. Instead, it empowers the network to comprehend the concept of velocity, enabling the conversion of offsets into relative time shifts. Additionally, unlike the sequential processing of one station at a time in the Taup algorithm, our module simultaneously calculates the time shifts for multiple stations associated with a single event. In essence, our association module can be considered a computationally efficient 3D Taup algorithm that operates without requiring a velocity model.
To evaluate the estimation accuracy of the arrival time using this module, we applied the estimated time shifts to align different stations (Fig. S4). 
Because the multi-station association module can accurately estimate the arrival time, 
the original waveforms from all stations were aligned accordingly. 

\begin{figure}[t!]%
\centering
\includegraphics[width=0.8\textwidth]{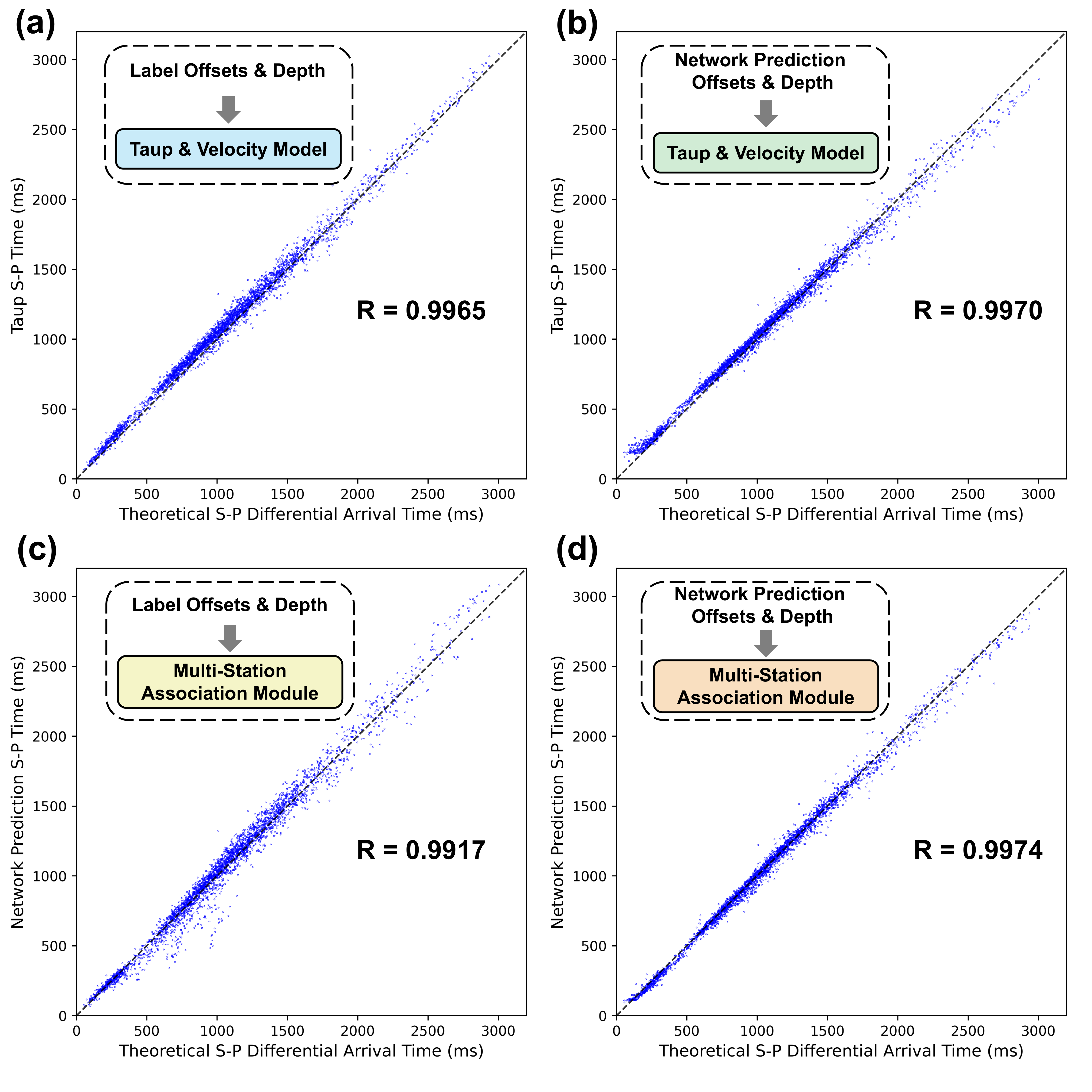}
\caption{Comparison of the P/S-wave arrival time estimation using different methods.
(\textbf{A} and \textbf{B}) The crossplots of the S-P differential arrival times computed by the TauP model with the input of offsets and depths from manual labels and network predications, respectively. The x-axis represents the manually picked S-P differential arrival times. The y-axis represents the S-P differential arrival times obtained by different methods. (\textbf{C} and \textbf{D}) The crossplots of the S-P differential arrival times predicted by our multi-station association module, and they are consistent with those predicted by the TauP model. This indicates that the multi-station association module, detached from the entire trained PLAN model, works physically reasonable compared to the commonly used TauP model.
}\label{fig7}
\end{figure}

To further evaluate the estimation accuracy of the arrival time using this module, we employed the TauP algorithm based on the PREM model~\citep{dziewonski1981preliminary} for comparison (Fig. \ref{fig7}). 
We also calculated the correlation coefficient (R) between the output of each method and the manually picked P/S-wave time differences. 
During the training process, this module used the offsets and depths obtained from the earthquake location module as inputs. 
Therefore, inputting the predicted offsets and depths into this module (Fig. \ref{fig7}D) 
could yield better P/S-wave time differences than inputting the labeled offsets and depths (Fig. \ref{fig7}C). 
The multi-station association module with manually labeled offsets and depths yielded less consistent results than the TauP algorithm. 
This discrepancy may not be solely attributable to errors in the deep learning estimation. Label inaccuracies may have also contributed to this outcome.
This assertion was supported by the observation that using the neural network output as the input for the TauP algorithm resulted in greater correlation coefficients than when label was employed (Fig. \ref{fig7}A-\ref{fig7}B). 
Among all the evaluated methods, the estimation results in Fig. \ref{fig7}D 
show the highest correlation coefficients.  
Generally, the multi-station association module and the TauP algorithm based on the PREM model have the same level of accuracy in calculating P/S-wave time differences.

In addition, the out-of-distribution data is utilized to assess the network's temporal generalization ability and application potential.
We applied PLAN and PhaseNet to pick the aftershock sequence of the 2022 M 7.4 Fukushima earthquake in Japan (Fig. S5).  
Traditional deep learning-based phase picking methods, such as PhaseNet, typically focus on picking seismic phases at a single station and perform pick and association processes step-by-step. 
In such a sequentially independent processing flow, the accuracy of phase picking determines the precision of subsequent earthquake association and location. Additionally, the consistency constraints among multiple stations for the same seismic event in the subsequent association process cannot be fed back to the previous step of phase picking to improve its accuracy. Without the consistency constraints among multiple stations, the phase picking station-by-station may be sensitive to noise and fail to pick reasonably pick multiple P/S waves at stations with short epicental distances as shown in Fig. S5A-S5B.
In contrast, PLAN is a simultaneous multi-station and multi-task processing approach that allows the aggregation of information from all stations and tend to pick P/S waves that can be consistently associated to the same events recorded by all the stations, leading to more reliable picking results. Furthermore, during the network training process, the picking, association, and location modules in the PLAN network mutually constrain and provide feedback to each other. As a result, a coherent and adaptive organic system can be achieved, simultaneously accomplishing these three tasks in a cohesive and integrated manner.

Some limitations remain in our method, particularly in handling continuous waveform data. This is mainly related to the construction of our training dataset and the corresponding training strategy. In our training data, although we allow for varying numbers of stations and missing phase picking labels for some stations, we assume that each training sample contains only one earthquake event. We overlook the scenario where the sample data does not contain an earthquake event (only noise), and when the data contains multiple events, we only provide labels for the main event. These considerations simplify the process of constructing our training samples but limit the diversity of the samples and the ability of the trained network model to handle continuous waveform data. However, we believe that these limitations can be addressed by using a more diverse and comprehensive training dataset.

In summary, we present a novel all-in-one multi-task multi-station system called PLAN for earthquake monitoring, which is capable of simultaneous phase picking, phase association, and earthquake location. 
Unlike current CNN-based methods that perform phase picking station-by-station, phase association and location separately, 
our proposed GNN-based multi-station multi-task system best utilizes the inherent inter-task and inter-station constraints. 
The multi-station association module estimates the phase shift and improves the robustness and accuracy of the phase association process. 
Eventually, the resulting offsets and depth enables accurate event localization. Our method demonstrates the need to factor mutual constraints among tasks and stations into next-generation earthquake monitoring systems. 

\section*{Methods}\label{sec4}
\subsection*{Graph based neural network}\label{subsec5}

Several studies have shown that GNNs have the potential to deal with irregularly spaced stations for 
phase association and event localization~\citep{mcbrearty2019earthquake,van2020automated,yano2021graph,mcbrearty2022earthquake1,
mcbrearty2023earthquake,zhang2022spatio,bilal2022early}. 
Here, we build a graph-based network (Fig. \ref{fig1}) for mult-station earthquake monitoring.
To utilize the GNN, we first need to change the data from
the matrix format to the graph format and employ a graph-based representation 
of the stations, where each station is represented as a node in the graph
and the three-channel data and the station location are used as the features of each node.
In contrast to the current single-station processing methods~\citep{ross2018generalized,ross2018p,zhu19phasenet,mousavi2019cred,
zhu2019deep,pardo2019seismic,wang2019deep,liu2020rapid,mousavi2020earthquake,
zhu2022ustc}, which treat each three-channel data as an individual input sample, our approach inputs all the three-channel data received from multiple stations per event as a single sample. 
This allows for efficient aggregation of information from multiple stations during network training. 
As a result, 
the features of different stations could be effectively integrated using GNNs during the aggregation process.

In this study, we have evaluated various graph aggregation methods, including
GCN~\citep{kipf2016semi}, GraphSAGE~\citep{hamilton2017inductive}, GAT~\citep{velivckovic2017graph},
GATv2~\citep{brody2021attentive}, and TransformGCONV~\citep{shi2020masked}. 
Through this evaluation, we have determined that TransformGCONV, which is based on attention
mechanism~\citep{vaswani2017attention}, is the most suitable module for the proposed PLAN. 
The message aggregation of TransformGCONV could be represented as:

\begin{equation}
\mathbf{x}_{i}^{\prime}=\mathbf{W}_1 \mathbf{x}_i+\sum_{j \in \mathcal{N}(i)} 
\alpha_{i,j} \mathbf{W}_2 \mathbf{x}_j,
\end{equation}
where $\mathbf{x}_{i}^{\prime}$ represents the aggregated features at the source node, and $\mathbf{x}_{i}$ and $\mathbf{x}_{j}$ represent the features of the source and distant nodes before aggregation, respectively.
$\mathbf{W}_1$ and $\mathbf{W}_2$ are the trainable matrices. 
In addition, the attention coefficients $\alpha_{i,j}$ are computed via dot-product attention as follows:

\begin{equation}
\alpha_{i, j}={softmax}\left(\frac{\left(\mathbf{W}_3 \mathbf{x}_i\right)^{\top}\left(\mathbf{W}_4 \mathbf{x}_j\right)}{\sqrt{d}}\right),
\end{equation}
where $\mathbf{W}_3$ and $\mathbf{W}_4$ are the trainable matrices.
Similar to the attention mechanism\citep{vaswani2017attention}, the source feature $\mathbf{x}_{i}$ 
and distant feature $\mathbf{x}_{j}$ are transformed into query vector
and key vector, respectively, using $\mathbf{W}_3$ and $\mathbf{W}_4$.
Compared to other graph aggregation methods, the use of the attention-based mechanism (equation 1) in TransformGCONV allows for 
a more fine-grained representation of the relationship between different stations,
thereby improving the accuracy and efficiency of the proposed method.

\subsection*{Network Architecture}\label{subsec6}

Here, we design a multi-station multi-task network for simultaneous phase picking, association, and location. 
The network (Fig. S1) comprises four components: a waveform feature extraction module, an earthquake location module, a multi-station association module, and a physics-informed multi-station phase-picking module.
Similar to previous deep-learning-based phase-picking approaches~\citep{zhu19phasenet,mousavi2020earthquake,zhu2022ustc,zhu2022end}, we design an encoder to extract waveform features and a decoder to produce phase-picking results.
However, to address the multi-station phase-picking problem, we introduce the GNN-based
TransformGCONV for aggregating features from multiple stations.

Because aligned waveform features are easily used and aggregated in GNNs for multi-station phase picking, we do not employ it in the waveform feature extraction module (Fig. S1A), where the features are relatively shifted in time.  
Although we use a U-shape neural network for feature extraction to solve the phase-picking problem, it could be replaced with other single-station-based phase-picking networks, such as EQTransformer.
No matter what type of network architecture is used, the features extracted from the middle of the network are input into the earthquake location module, and the structure of the final few layers of the network are modified for the purpose of multi-station phase picking.
Additionally, the kernel size of all convolutional layers in the waveform feature extraction network is set to 7.

For the earthquake location module (Fig. S1B), we first extract features from the normalized coordinate information of the stations within the range of [0,1] through two fully connected layers (3-48-96). 
Simultaneously, the waveform features extracted from each station are
further processed through several convolutional layers and then flattened.
Subsequently, the position and waveform features are concatenated and passed through two fully connected layers (192-192-96). 
This fuses the position and waveform features at each station. 
The fused features are further aggregated among multiple stations by several GNN layers to predict the 
offsets of each station with respect to the event and its depth. 
Because there is only one depth parameter for each sample, we add a global average pooling before the output. 
In summary, this module allows the integration of both location and waveform information into the feature extraction process, which is crucial for accurate event localization.

Finally, in the physics-informed multi-station phase-picking module (Fig. S1D), 
we incorporate physics-motivated constraints of time alignment among waveforms corresponding to the same earthquake event. 
We first utilize a mulit-station association module (Fig. S1C) to calculate the relative alignment shifts between stations using the estimated offsets and the depth of the event. 
We then use the shifts to align the waveforms to a common time standard and aggregate the features across multiple stations in the phase-picking module. 
Subsequently, the aggregated features at each station are unaligned and fed to two layers of convolution to yield final P/S-wave picking results. 
This process leverages the physical information of the event location to improve the robustness and accuracy of the multi-station phase picking.

\subsection*{Loss function and training details}\label{subsec7}

Our multi-task learning network model has three output results 
corresponding to phase picking, phase association, and earthquake localization. 
To train the model, we define three different loss functions 
for these three different tasks. 
For phase picking, instead of using the commonly used cross-entropy, 
we choose the mean square error (MSE) as the loss function, 
which is suitable for training in multi-task problems. 
To estimate offsets and depth, which is similar to event localization, 
we also use MSE as the loss function, as suggested by previous studies
\citep{van2020automated,zhang2022spatio}. 
Finally, to calculate the P/S-wave shift, 
we define the loss function as follows:

\begin{equation}
\mathcal{L}_{\Delta p}=\sum_{i=1}^n\mid {CTime}_{p}-\left({label}_{p_i}+\Delta t_{p_i}\right)\mid,
\end{equation}
\begin{equation}
\mathcal{L}_{\Delta s}=\sum_{i=1}^n\mid {CTime}_{s}-\left({label}_{s_i}+\Delta t_{s_i}\right)\mid,
\end{equation}

where $ {CTime}_{p}$ and $ {CTime}_{s}$ represents the 
reference times where the P- and S-wave picks are aligned, respectively. 
Moreover, $ {label}_{p_i}$ or 
$ {label}_{s_i}$ represents the manually picked P/S-wave arrival time 
for each station, and $\Delta t$ represents the predicted P/S-wave shift.
Finally, we combine the three types of loss functions to form the overall 
objective function:
\begin{equation}
L_{ {total }}=\lambda_1 \mathcal{L}_{ {picking-p }}+\lambda_1\mathcal{L}_{ {picking-s }}+\lambda_2\mathcal{L}_{\Delta p}+\lambda_2\mathcal{L}_{\Delta s}+\lambda_3\mathcal{L}_{ {offset}}+\lambda_3\mathcal{L}_{ {depth }}.
\end{equation}
Here, we set the coefficients $\lambda_1$, $\lambda_2$, 
and $\lambda_3$ to 1.

During the training process, the model was optimized using the ADAM~\citep{kingma2014adam} method with an initial learning rate of 0.001, which is gradually decreased with a decay rate of 0.9 every 100 epochs. 
To enhance the training efficiency, we randomly selected 2048 events from the training set for each epoch, rather than using the entire data. 
The model was trained for a total of 2000 epochs with a batch size of 16, 
and the training process required approximately 24 h using 1 NVIDIA Tesla A100 GPU.

\subsection*{Evaluation metrics}\label{subsec7}

In previous studies \citep{mousavi2020earthquake,zhu2022end}, true positive phase picks were defined as those within 0.5 s of the predicted pick. The rest were counted as false positives.
Nevertheless, owing to potential errors in the labels of the dataset, such statistical results based on a single threshold may not be reliable.
Thus, to better evaluate the performance of algorithms, we introduce new metrics, mPrecision, mRecall, and mF1, which are calculated using multiple thresholds, following previous research\citep{zheng2022clrnet}. The metrics are defined as:

\begin{equation}
\mathrm{mPrecision}=(\mathrm{Precision} @ 11+\mathrm{Precision} @ 12+\cdots+\mathrm{Precision} 1 @ 50) / 40,
\end{equation}
\begin{equation}
\mathrm{mRecall}=(\mathrm{Recall} @ 11+\mathrm{Recall} @ 12+\cdots+\mathrm{Recall} 1 @ 50) / 40,
\end{equation}
\begin{equation}
\mathrm{mF} 1=(\mathrm{F} 1 @ 11+\mathrm{F} 1 @ 12+\cdots+\mathrm{F} 1 @ 50) / 40.
\end{equation}
where x@11, x@12, $\cdots$ , x@50 are Precision, Recall, or F1 metrics when the thresholds are 11, 12, $\cdots$, 
50 samples (corresponding to 0.11 s, 0.12 s, $\cdots$, 0.5 s of time), respectively.
These metrics, mPrecision, mRecall, and mF1 reward detectors with better picking results and, therefore, can more reasonably or fairly assess the performance of the different methods. 


\bibliography{arxiv_science.bib}

\bibliographystyle{plainnat}

%





\section*{Supplementary Information}\label{sec7}

\setcounter{figure}{0}
\renewcommand{\thefigure}{S\arabic{figure}}
\begin{figure}[h!]%
\centering
\includegraphics[width=\textwidth]{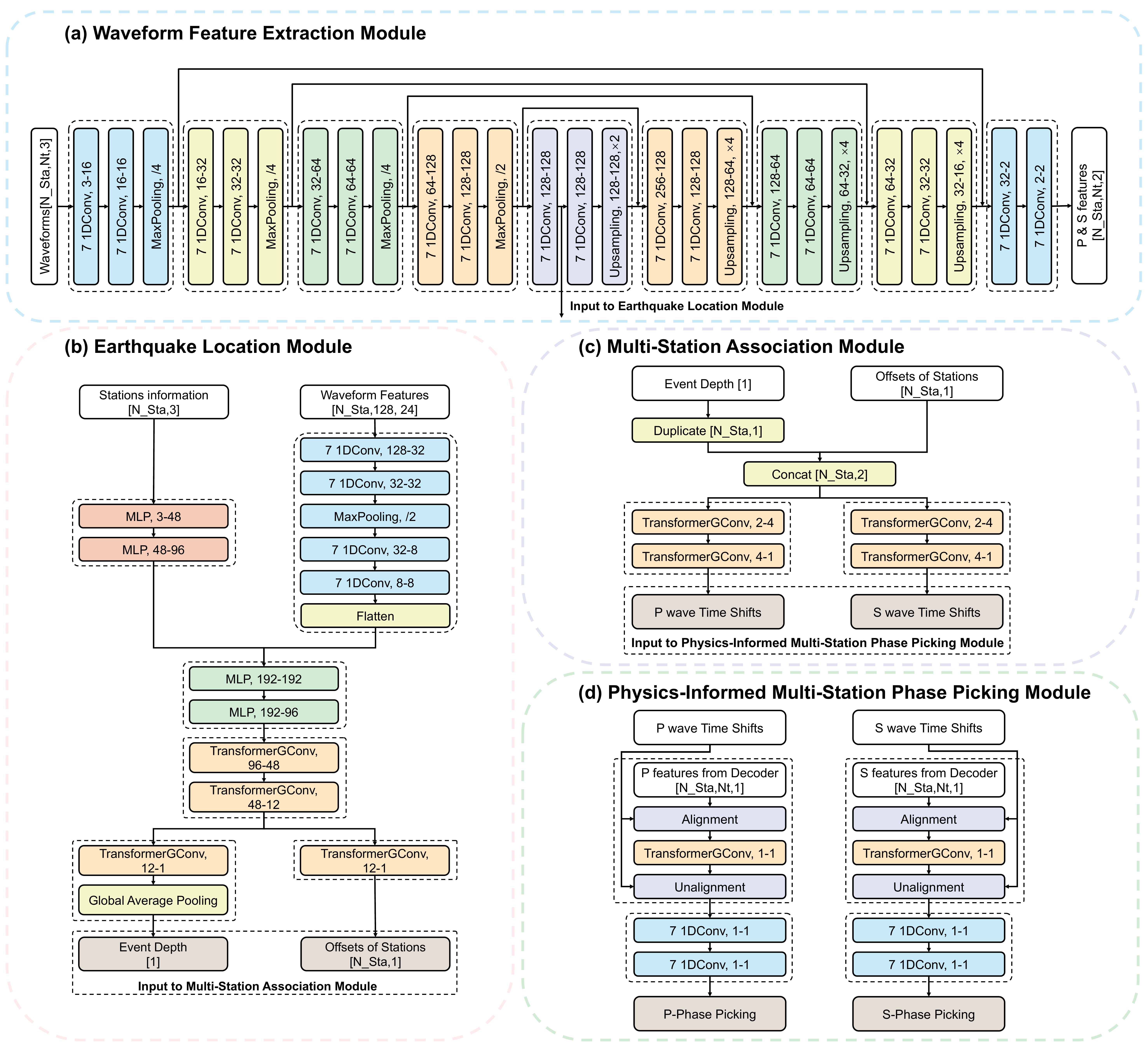}
\caption{Network architecture of PLAN. 
\textbf{a} waveform feature extraction encoder and decoder extract features from raw seismic waveforms using a simplified 1D U-net. These features, along with information about the locations of the stations, are input into the earthquake location module \textbf{b}, which is a two-branch module. It extracts and combines waveform and location information for each station and calculates the offsets of the stations relative to the event and the depth of the event. \textbf{c} multi-station association module then uses the offsets and depth information to estimate the relative time shifts between different stations which are further used to temporally align the waveform features from different stations. \textbf{d} finally, the phase-picking module aggregates features (among multiple stations) in the aligned space and then predicts P/S-wave phase arrivals in the original space. Furthermore, in 1DConv blocks, the preceding number indicates the size of the convolutional kernel, while the subsequent numbers respectively denote the input and output channels. Regarding the TransformerGConv Blocks, the following numbers correspondingly represent the input and output features.
}\label{fig:supp_figure1}
\end{figure}

\begin{figure}[htb!]%
\centering
\includegraphics[width=\textwidth]{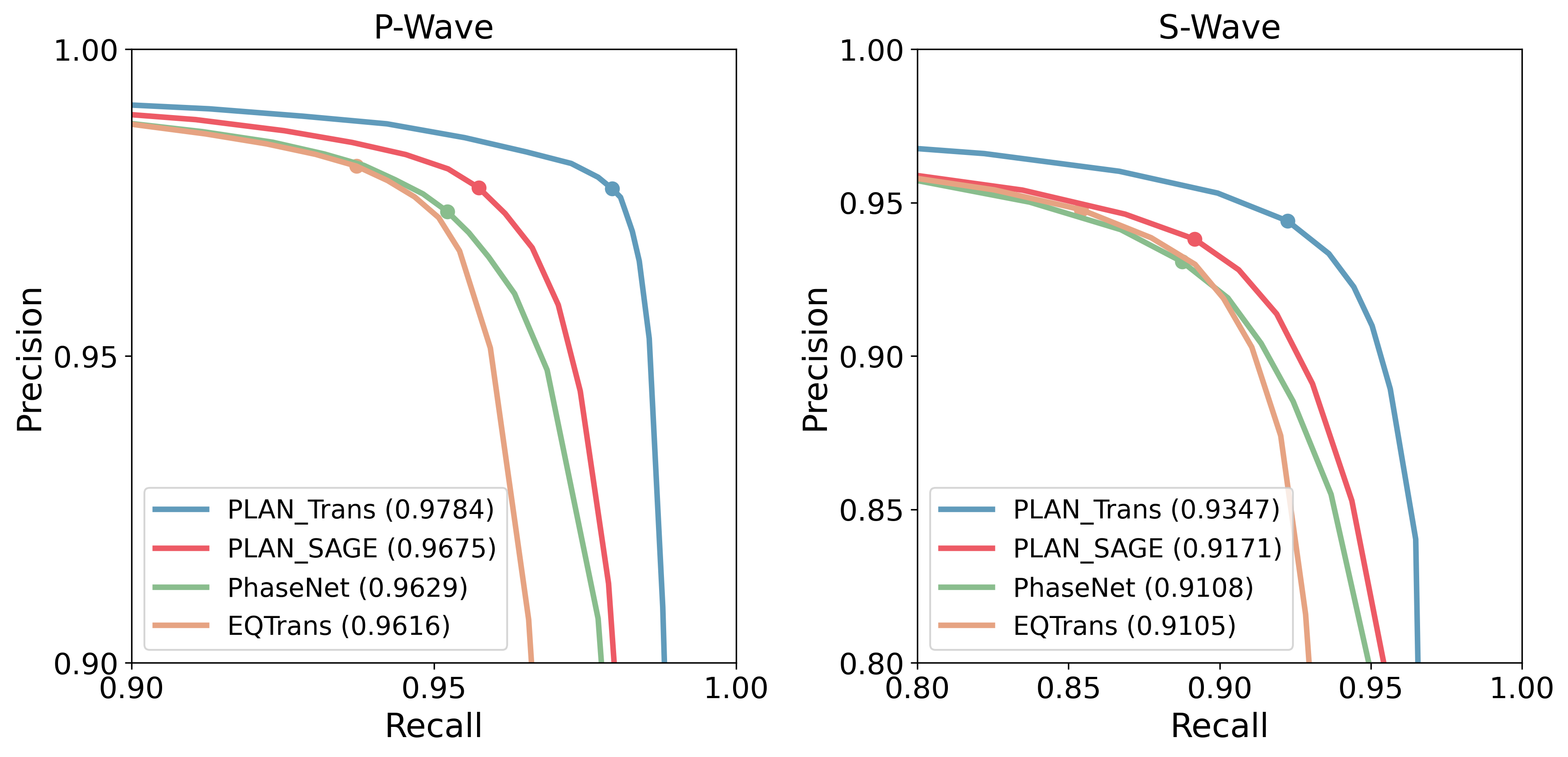}
\caption{Comparison of the precision-recall curves of picking results in the Ridgecrest region. True positive phase picks were defined as those within 0.5 s of the predicted pick. The numbers in legend represent the maximum F1 scores of each model.
}\label{fig:supp_figure2}
\end{figure}

\begin{figure}[htb!]%
\centering
\includegraphics[width=\textwidth]{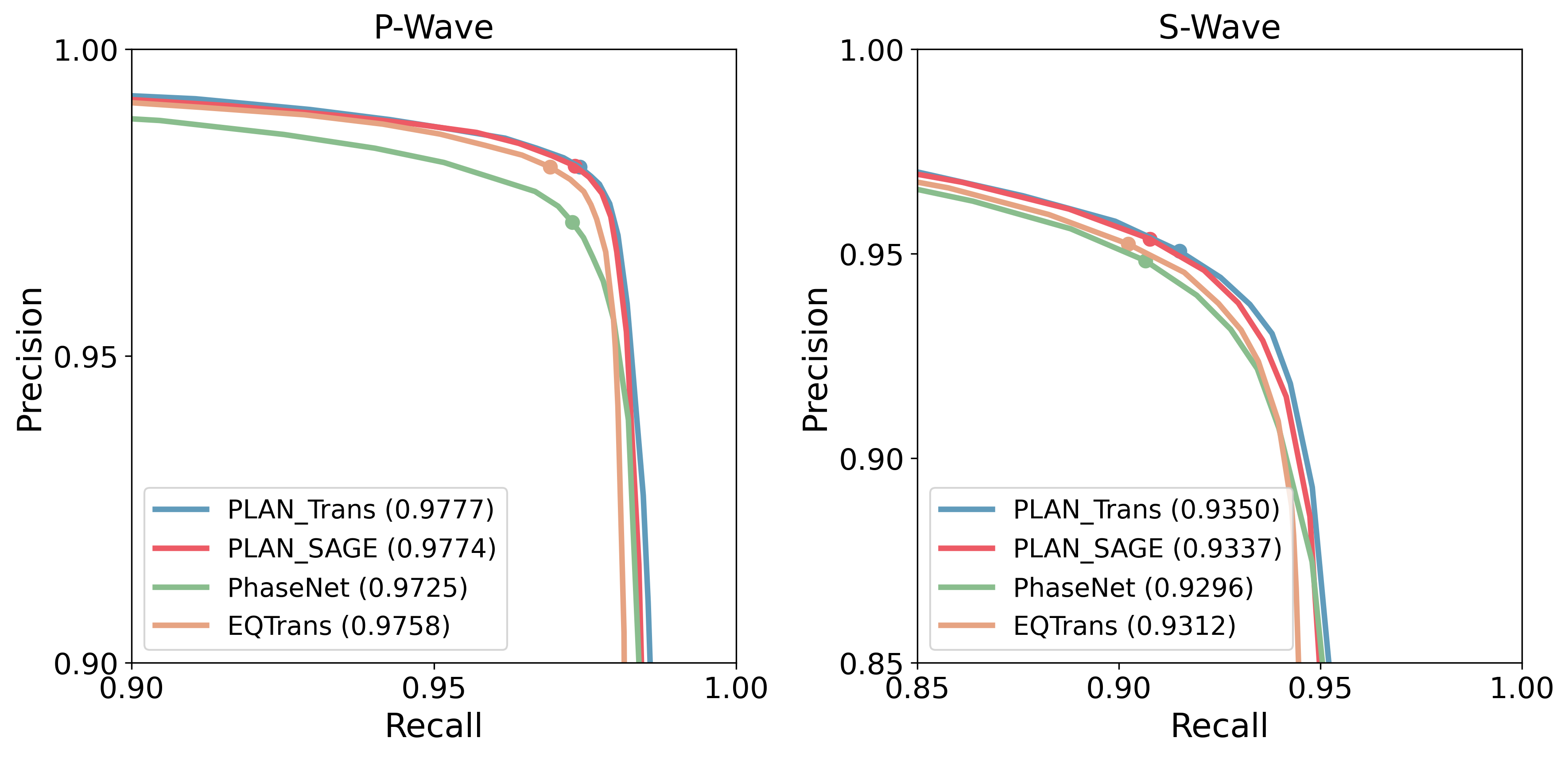}
\caption{Comparison of the precision-recall curves of picking results in Japan. True positive phase picks were defined as those within 0.5 s of the predicted pick. The numbers in legend represent the maximum F1 scores of each model.
}\label{fig:supp_figure2}
\end{figure}

\begin{figure}[htb!]%
\centering
\includegraphics[width=\textwidth]{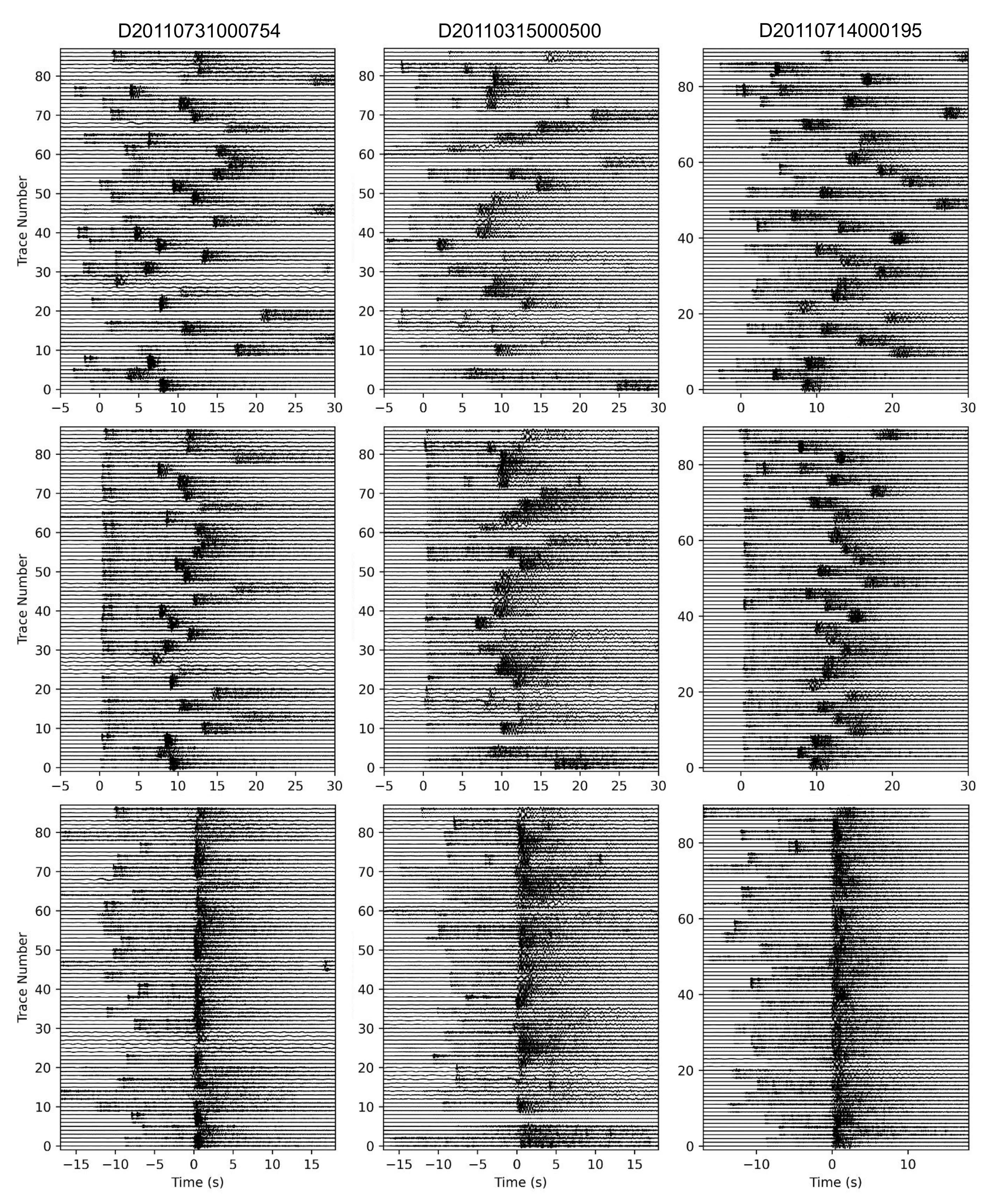}
\caption{Example of raw waveforms (first line) and aligning raw waveforms of P waves (second line) and S waves (third line) using the time shifts predicted by the multi-station association module. The numbers on the top of the figures represent the identification number of the events in the Japan Meteorological Agency (JMA) catalog. The alignment results indicate that the multi-station association module can accurately estimate the time shifts and further utilize it to align the waveform features at the same time temporal position.
}\label{fig:supp_figure2}
\end{figure}

\begin{figure}[htb!]%
\centering
\includegraphics[width=\textwidth]{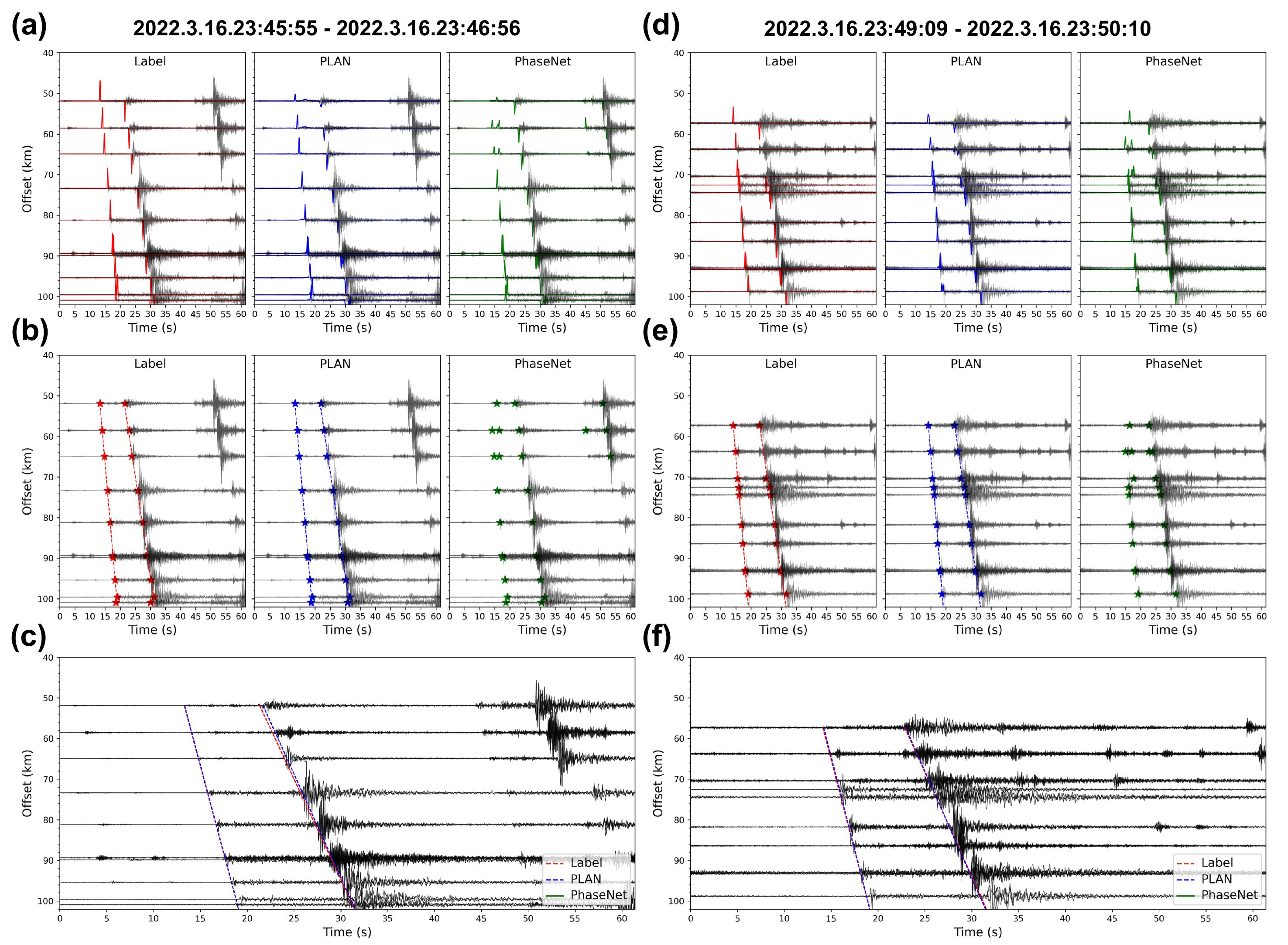}
\caption{Comparison of waveform picking results for the aftershock sequence of the 2022 M 7.4 Fukushima earthquake in the Japan using PLAN and PhaseNet. The red, blue, and green markers represent the manual labels, PLAN results, and PhaseNet results. The lines in \textbf{b}, \textbf{c}, \textbf{e} and \textbf{f} are obtained by linear fitting based on the picking results. PhaseNet picks may become mixed in near-offset stations, making it challenging to obtain a reliable result without phase association. In contrast, due to its nature as a multi-task method that simultaneously performs phase picking and phase association, PLAN exhibits stronger consistency in picking results across stations.
}\label{fig:supp_figure3}
\end{figure}

\clearpage

\end{document}